\PassOptionsToPackage{pdfpagelabels=false}{hyperref}
\documentclass[fleqn,usenatbib,usedcolumn]{mnras}
\usepackage[british]{babel}             
\usepackage{txfonts}                  

\usepackage{graphicx}	
\usepackage{hyperref}	

\newcommand{\rgzwat}{RGZ J082312.9+033301}	
\newcommand{\newcluster}{Matorny-Terentev}	         

\DeclareRobustCommand{\ion}[2]{%
\relax\ifmmode
\ifx\testbx\f@series
{\mathbf{#1\,\mathsc{#2}}}\else
{\mathrm{#1\,\mathsc{#2}}}\fi
\else\textup{#1\,{\mdseries\textsc{#2}}}%
\fi}

\title[Discovery of a giant WAT]{Radio Galaxy Zoo: discovery of a poor cluster through a giant wide-angle tail radio galaxy}

\author[J.~K.~Banfield et al.]{J.K. Banfield$^{1,2}$\thanks{E-mail: \href{mailto:julie.banfield@anu.edu.au}{julie.banfield@anu.edu.au} (JKB); \href{mailto:heinz@astro.ugto.mx}{heinz@astro.ugto.mx} (HA)},
H. Andernach$^3$, A.D. Kapi\'{n}ska$^{4,2}$, L. Rudnick$^{5}$, M.J. Hardcastle$^{6}$, \newauthor 
G. Cotter${^7}$, S. Vaughan$^{7}$, T.W. Jones$^{5}$, I. Heywood$^{8,9}$, J.D. Wing$^{10}$, O.I. Wong$^{4,2}$, \newauthor
T. Matorny$^{11}$, I.A. Terentev$^{11}$, \'A.~R.~L\'opez-S\'anchez$^{12,13}$, R.P. Norris$^{8,14}$, N. Seymour$^{15}$, \newauthor
S.S. Shabala$^{16}$, K.W. Willett$^{5}$
\\
$^{1}$Research School of Astronomy and Astrophysics, Australian National University, Canberra ACT 2611, Australia\\
$^{2}$ARC Centre of Excellence for All-Sky Astrophysics (CAASTRO)\\
$^{3}$Departamento de Astronom\'ia, DCNE, Universidad de Guanajuato, Apdo.\ Postal 144, CP 36000, Guanajuato, Gto., Mexico\\
$^{4}$International Centre for Radio Astronomy Research-M468, The University of Western Australia, 35 Stirling Hwy, Crawley, WA 6009, Australia\\
$^{5}$School of Physics and Astronomy, University of Minnesota, 116 Church St. SE, Minneapolis, MN 55455, USA\\
$^{6}$Centre for Astrophysics Research, School of Physics, Astronomy and Mathematics, University of Hertfordshire, College Lane, Hatfield AL10 9AB, UK\\
$^{7}$Oxford Astrophysics, Denys Wilkinson Building, Keble Road, Oxford OX1 3RH, UK\\
$^{8}$CSIRO Astronomy and Space Science, Australia Telescope National Facility, PO Box 76, Epping, NSW 1710, Australia\\
$^{9}$Department of Physics and Electronics, Rhodes University, PO Box 94, Grahamstown 6140, South Africa\\
$^{10}$Harvard-Smithsonian Center for Astrophysics, 60 Garden St., Cambridge, MA 02138 USA\\
$^{11}$Zooniverse Citizen Scientist, c/o Oxford Astrophysics, Denys Wilkinson Building, Keble Road, Oxford OX1 3RH, UK\\
$^{12}$Australian Astronomical Observatory, PO Box 915, North Ryde, NSW 1670, Australia\\
$^{13}$Department of Physics and Astronomy, Macquarie University, NSW 2109, Australia\\
$^{14}$Western Sydney University, Locked Bag 1797, Penrith South, NSW 1797, Australia\\
$^{15}$International Centre for Radio Astronomy Research, Curtin University, Perth,  WA 6102, Australia\\
$^{16}$School of Physical Sciences, University of Tasmania, Private Bag 37, Hobart, Tasmania 7001, Australia
}

\date{Accepted 2016 May 2. Received 2016 March 9}

\pubyear{2016}

\begin{document}
\label{firstpage}
\pagerange{\pageref{firstpage}--\pageref{lastpage}}
\maketitle

\begin{abstract}
We have discovered a previously unreported poor cluster of galaxies (RGZ-CL J0823.2+0333) through an unusual giant wide-angle tail radio galaxy found in the Radio Galaxy Zoo project. We obtained a spectroscopic redshift of $z=0.0897$ for the E0-type host galaxy, 2MASX J08231289+0333016, leading to M$_r = -22.6$ and a $1.4\,$GHz radio luminosity density of  $L_{\rm 1.4} = 5.5\times10^{24}$ W Hz$^{-1}$. These radio and optical luminosities are typical for wide-angle tailed radio galaxies near the borderline between Fanaroff-Riley (FR) classes I and II. The projected largest angular size of $\approx8\arcmin$ corresponds to $800\,$kpc and the full length of the source along the curved jets/trails is $1.1\,$Mpc in projection. X-ray data from the {\it XMM-Newton} archive yield an upper limit on the X-ray luminosity of the thermal emission surrounding \rgzwat\,at $1.2-2.6\times10^{43}$ erg s$^{-1}$ for assumed intra-cluster medium temperatures of $1.0-5.0\,$keV. Our analysis of the environment surrounding \rgzwat\ indicates that \rgzwat\,lies within a poor cluster. The observed radio morphology suggests that (a) the host galaxy is moving at a significant velocity with respect to an ambient medium like that of at least a poor cluster, and that (b) the source may have had two ignition events of the active galactic nucleus with $10^7\,$yrs in between.  This reinforces the idea that an association between \rgzwat\,and the newly discovered poor cluster exists. 
\end{abstract}



\clearpage
\section{Introduction}
High-resolution radio surveys performed over the past decades have shown the wide variety of radio morphologies of galaxies illustrating the complexity of the underlying physics.  The majority of radio sources have compact morphology  \citep{Shabala2008,Sadler2014} while the extended radio-loud sources tend to be Fanaroff and Riley (FR) type I and II \citep{Fanaroff1974}.  However, there are some extended radio-loud sources that do not fit the standard FRI or FRII classification.  

The division of tailed radio galaxies into narrow-angle tails (NAT, head) and wide-angle tails (WATs) was introduced by \citet{Owen1976} and \citet{Rudnick1976}. Tailed radio galaxies have provided evidence that in both dense and sparse environments, the bending and distortions of radio galaxies are the result of motions with respect to the thermal plasma. WATs and straight FRI sources are often associated with the brightest galaxies in clusters (BCGs).  Their radio morphologies reflect both the initial jet momentum and the mild effects of motions \citep[e.g., ][have shown that a large number of BCGs have a significant peculiar velocity compared to their cluster mean]{Coziol2009} and pressure gradients in the intracluster medium \citep[ICM; ][]{Pinkney1994,Giacintucci2009}.   

WATs generally have C-shaped morphologies and have radio luminosities near the FRI and FRII luminosity transition \citep{Owen1994}. WATs are found in both merging and relaxed clusters at, or near, the centre and display highly collimated jets.  Early models of WATs suggested that ICM ram pressure resulting from velocities $> 1000\,$km s$^{-1}$ was required to produce the observed bends \citep{Eilek1984,ODonoghue1993} but models using light jets by \citet{Sakelliou1996}, \citet{Hardcastle2005}, \citet{Jetha2006} and \citet{Mendygral2012} show that bulk velocities around $100\,$km s$^{-1}$ are sufficient to produce WATs.

FR II radio sources have their size and shape dominated by the momentum of the overpressured jet. As the jet expands it develops a cocoon and a more collimated jet flows out to power the hot spots.  The jets in FRI-NATs undergo a complete disruption, after which they are carried back by external motions alone with no surrounding cocoon.  Intermediate between these extremes, FRI-WATs and straight FRIs have cocoons surrounding the outer portions of their jets \citep{Hardcastle1998,Katz-Stone1999}, so both jet momentum and motions in the surrounding thermal medium influence the subsequent flow.  

Questions remain unanswered from both the models and the observations.  Are any bright radio spots or knots in the jets powered by an impinging jet? Has the energy supply shut off, so that the knots are in the process of fading due to radiative losses and mechanical dissipation?  Is the jet outflow exposed to the external plasma or shielded by a stationary or slower-moving cocoon of relativistic plasma as observed around FRII and some FRI jets? 

The Radio Galaxy Zoo discovery of a giant WAT (\rgzwat) shows the power of using bent radio sources as tracers of clusters.  Upcoming wide-area radio surveys the Evolutionary Map of the Universe \citep[EMU; ][]{Norris2011}, the WODAN survey \citep{Rottgering2011}, and the deeper radio survey MeerKAT MIGHTEE \citep{Jarvis2012} are expected to detect over 100,000 bent radio sources \citep{Norris2011}.  Radio Galaxy Zoo will allow us to locate these bent radio sources and to investigate the physics that allows jets to be tightly collimated while undergoing significant bending.

The present paper is organized as follows.  Section \ref{sec:wat} describes the discovery of the WAT while in Section \ref{sec:dis} we discuss the implications with respect to environment, dynamics, and the central AGN.  Section \ref{sec:conc} presents our conclusions. Throughout this paper we adopt a $\Lambda$CDM cosmology of $\Omega_{M} = 0.3, \,\Omega_{\Lambda} = 0.7$ with a Hubble constant of $H_{\rm 0} = 70\,$km s$^{-1}$ Mpc$^{-1}$.  With $z=0.0897$, the luminosity distance is $D_{L}=410\,$ Mpc and the angular size distance is $D_{A}=345.3\,$Mpc giving a scale of $1.674\,$kpc arcsec$^{-1}$ \citep{Wright2006}. We define the radio spectral index as $S_{\nu} \propto \nu^{\alpha}$.
%
%
%
%
%
 \section{Wide Angle Tail RGZ J082312.9+033301}\label{sec:wat}
 \begin{figure}
	\includegraphics[width=\columnwidth]{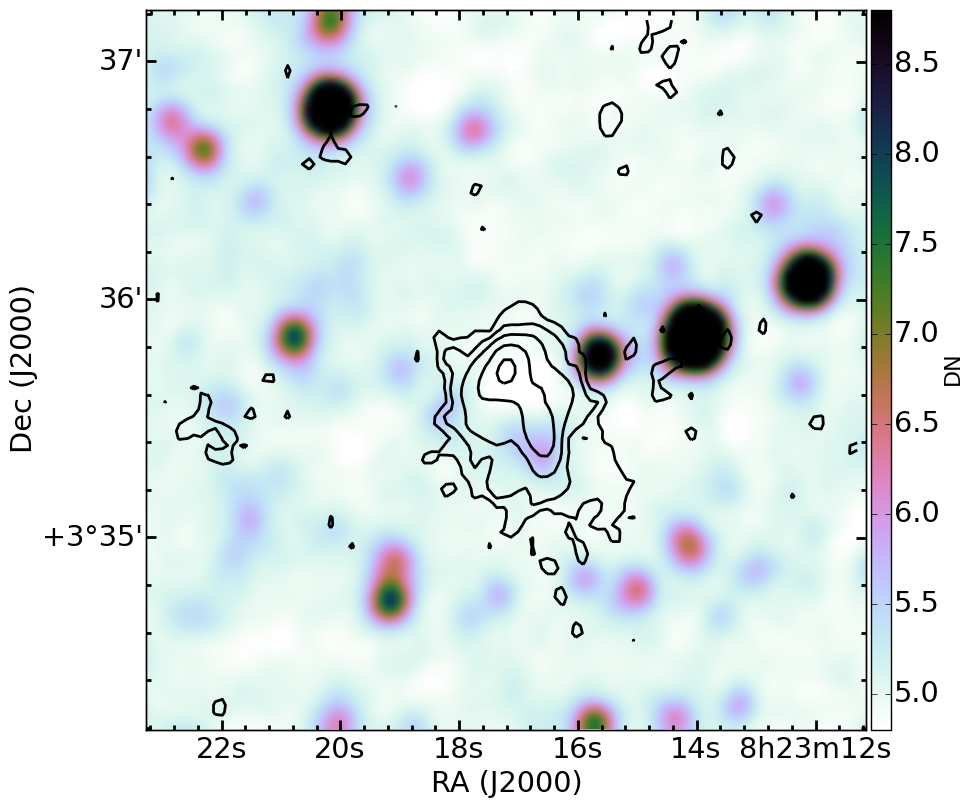}
    \caption{The Radio Galaxy Zoo $3 \times 3$ arcmin cutout of the FIRST radio data (black contours) with the WISE $3.4\,\mu$m image (grayscale) of the Radio Galaxy Zoo subject FIRSTJ082317.2+033542 that first indicated a possible larger object shown in Fig.~\ref{fig:2}.  The contours start at 2 times the local signal to noise ($1\sigma = 0.19\,$mJy beam$^{-1}$) and increase by multiples of 2. The background colour scheme comes from {\textsc CUBEHELIX} \citep{Green2011}. A colour version of the figure is available in the online version.}
    \label{fig:1}
\end{figure}
\begin{figure}
\includegraphics[width=\columnwidth]{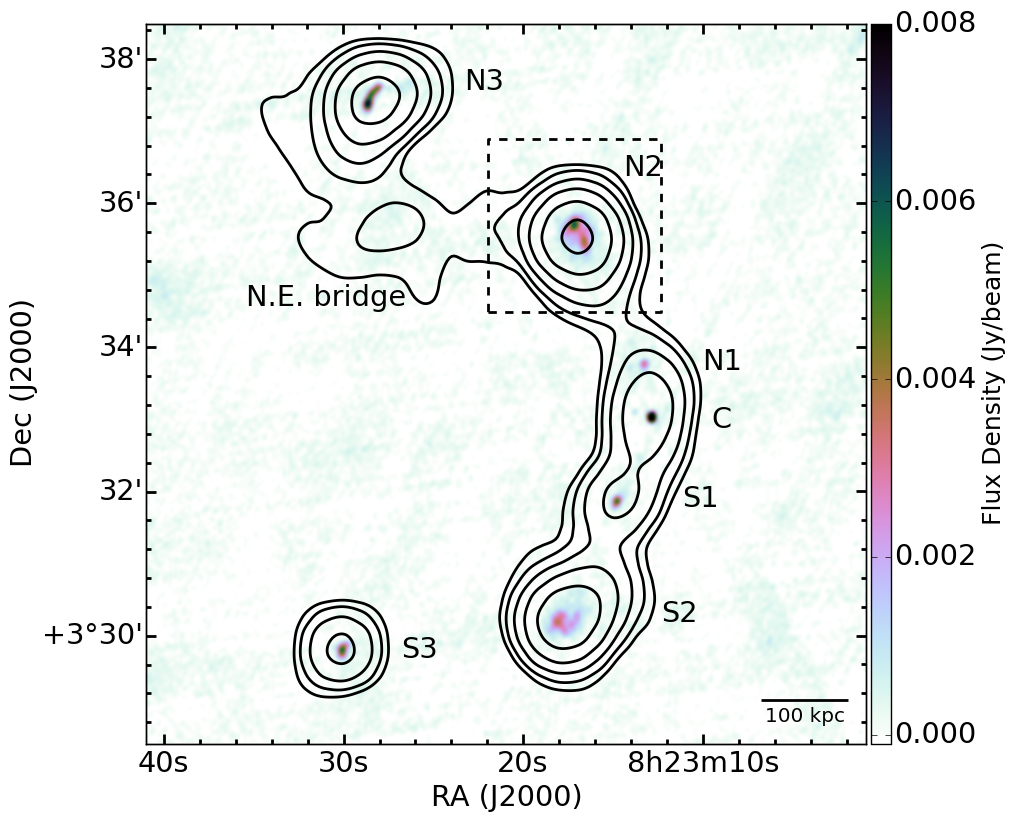}
\caption{The $1.4\,$GHz FIRST radio image (grayscale) with the $1.4\,$GHz NVSS contours shown in black. Contour levels begin at 2 times the local signal to noise level ($1\sigma = 0.83\,$mJy beam$^{-1}$) and increase by a factor of 2.  The components of RGZ J082321+033301 are labelled (clockwise from N.E.).  The dashed square indicates the area covered by the Radio Galaxy Zoo image from Fig.~\ref{fig:1}. A colour version of the figure is available in the online version.}
    \label{fig:2}
\end{figure}

\begin{table*}
 \centering
  \caption{The identification of the possible seven components
  of \rgzwat\ from the NVSS and FIRST catalogues shown in Fig.~\ref{fig:2}.}\label{tab:1}
 \begin{tabular}{lccccccl}
 \hline
Comp. & R.A. (J2000) & Dec. (J2000)  & FIRST I.D. & $S_{\rm 1.4\,FIRST}$ & NVSS I.D. & $S_{\rm 1.4\,NVSS}$ & Note\\
 & (degrees) & (degrees) & & (mJy) & & (mJy)\\
  \hline
  N3 & 125.869 & 3.623 & J082328+033722 & 15.38 & J082328+033724 & 56.4 & potential component\\
   & & & J082328+033733 & 11.56 \\
N2 & 125.824  & 3.593 & J082317+033534 & 12.28 & J082317+033530 & 87.5 &\\
 & & & J082316+033530 & 36.94 \\
 & & & J082317+033542 & 12.71\\
N1 & 125.806 & 3.563 & J082313+033345 & 6.28 & -- & -- & \\
C & 125.804 & 3.550 & J082312+033301 & 16.50 & J082313+033241 & 60.3 & \\
S1 & 125.812 & 3.531 & J082314+033151 & 10.08 & J082314+033047 & 3.8 & \\
S2 & 125.822 & 3.503 & J082317+033011 & 18.26 & J082317+033016 & 71.1 & \\
 & & & J082318+033012 & 21.33\\
 & & & J082317+033006 & 3.07\\
S3 &125.875 & 3.497 & J082330+032947 & 13.65 & J082330+032950 & 16.5 & unlikely component\\ 
\hline
\end{tabular}
\end{table*}
\begin{figure}
\includegraphics[width=\columnwidth]{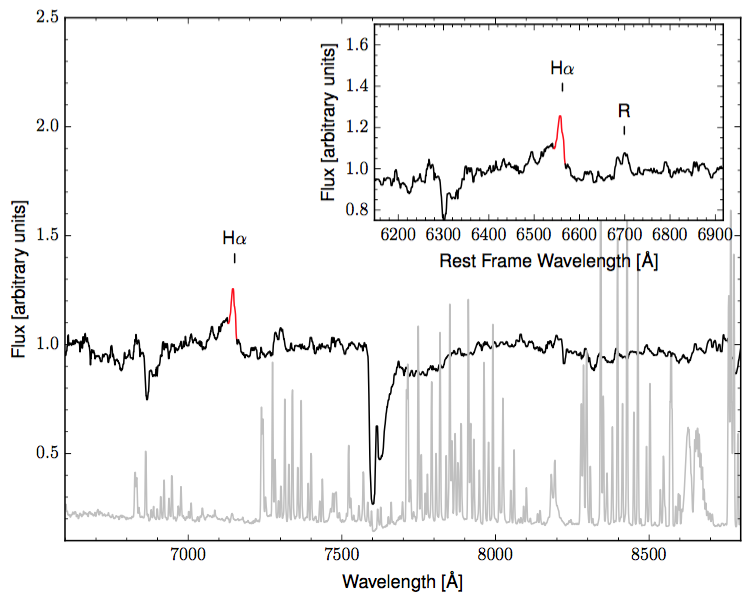}
\caption{The Oxford Swift IFU spectrum (black lines) of 2MASX J08231289+0333016 with the skylines (grey lines). The inset zooms into the region of the most important lines (H$\alpha$, [\ion{N}{ii}]~$\lambda$6583 or [\ion{S}{ii}]~$\lambda\lambda$6716,31). The redshifted H$\alpha$ line provides a redshift of $z = 0.0897\,\pm\,0.0001$. Note that the features near the expected position of the [\ion{S}{ii}] doublet (expected wavelength marked) is the skyline residuals.}
    \label{fig:3}
\end{figure}
\begin{figure*}
\includegraphics[scale=0.38]{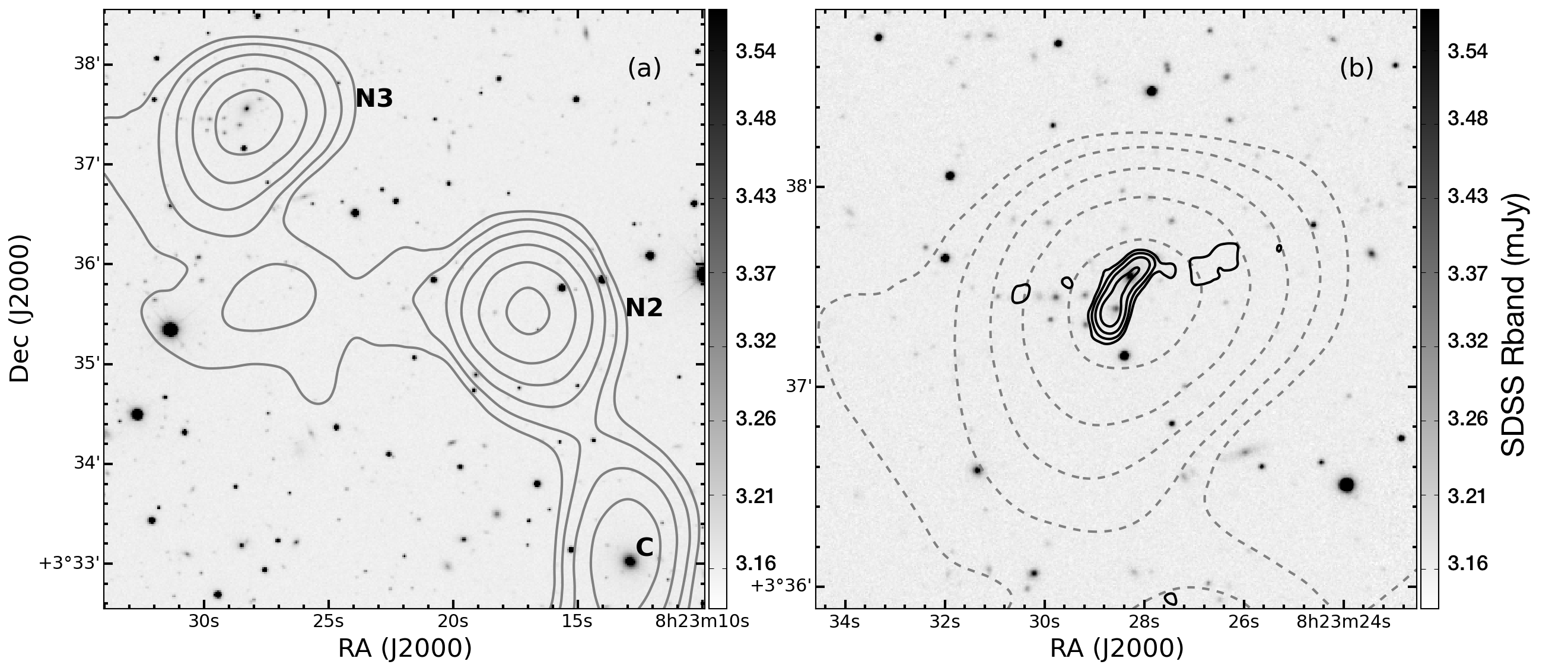}
\caption{(a) Enlargement of the N.E. bridge along with components C,
    N2 and N3 from Fig.~\ref{fig:2}.  The SDSS r-band image is shown
    in the inverted grayscale and the NVSS data is shown as
    contours, starting at 2 times the local noise ($1\sigma
    = 0.83\,$mJy beam$^{-1}$) and increasing by a factor of 2. (b) A
    $3 \times 3$-arcmin enlargement of N3 showing the possible alignment of N3 to a background cluster of galaxies at $z=0.2601$.  The NVSS data is shown as grey dashed contours, the FIRST data is shown as black solid contours, and the SDSS r-band image is shown in the inverted grayscale.  The FIRST contours start at 2 times the local noise ($1\sigma = 0.19\,$mJy beam$^{-1}$) and increasing by a factor of 2.}
    \label{fig:4}
\end{figure*}

The discovery of \rgzwat\,was made in the citizen science project Radio Galaxy Zoo\footnote{\href{http://radio.galaxyzoo.org}{http://radio.galaxyzoo.org}} \citep[RGZ;][]{Banfield2015}.  Radio Galaxy Zoo offers overlays of the $3.4\,\mu$m image from the {\it Wide-Field Infrared Survey Explorer} \citep[{\it WISE};][]{Wright2010} with the $1.4\,$GHz image from the Faint Images of the Radio Sky at Twenty Centimeters \citep[FIRST; ][]{White1997,Becker1995}.  The $3 \times 3$ arcmin$^2$ Radio Galaxy Zoo images are centred on the position of the radio sources listed as extended in the FIRST catalogue (version 14 March 2004) and then overlaid on the infrared image as we show in Fig.~\ref{fig:1}.  The radio images are illustrated with radio brightness contours, overlaid on a {\it WISE} $3.4\,\mu$m image in a heat scale. 

RGZ J082312+033301 was identified as an unusual object in December 2013 by two citizen scientists (T. Matorny \& I. Terentev) after examining the Radio Galaxy Zoo $3 \times 3$ arcmin$^2$ cutout (Fig.~\ref{fig:1}) of a section of the radio galaxy (N2 in Fig.~\ref{fig:2}).   Matorny first suggested that the radio emission pointed towards another object by way of the radio extension towards the S.W.. A further investigation of RGZ J082312+033301 by Terentev and Rudnick was completed in RadioTalk\footnote{\href{http://radiotalk.galalxyzoo.org}{http://radiotalk.galaxyzoo.org}} by examining the larger cutout of both the FIRST and {\it WISE} images along with images from the NRAO VLA Sky Survey \citep[NVSS; ][]{Condon1998} and the Sloan Digital Sky Survey (SDSS) Data Release 10 \citep{Ahn2014} and Data Release 12 \citep{Alam2015}.  It was then realized that the isolated component seen in Fig.~\ref{fig:1} was part of a much more extended radio source which could be classified as a WAT. 

In Fig.~\ref{fig:2} we show the FIRST image in a color scale and the NVSS image with contours.  We found that the core (C in Fig.~\ref{fig:2}) is coincident with 2MASX J08231289+0333016 (SDSS J082312.91+033301.3) and has 175 morphology votes from Galaxy Zoo 2 \citep{Willett2013} indicating that the host galaxy has morphological features consistent with type E0 with M$_r = -22.6$ and M$_V = -22.3$.  There is no spectrum of  SDSS J082312.91+033301.3 in SDSS DR12.  We used the Oxford Short Wavelength Integral Field Spectrograph for the Large Telescope (SWIFT) Integral Field Unit (IFU) spectrograph \citep{Thatte2010} on the Palomar 5-m Hale telescope to obtain an optical spectrum of SDSS J082312.91+033301.3 on the night of 2013 December 29 UT.  The target was observed with the large (0.235 arcsec) plate scale in natural seeing. We took two 300s exposures, offset from each other by 40 arcsec along the long axis of the detector to allow background subtraction.  We present the one-dimensional spectrum, extracted from a 7 arcec diameter circular aperture, in Fig.~\ref{fig:3}. 

Only one secure non-telluric feature is detected in this spectrum: a narrow line centered at 7151.18\AA. We identify the line as H$\alpha$ (rest wavelength of 6562.81\AA), providing a redshift of $z = 0.0897\,\pm\,0.0001$. The uncertainty in the redshift is dominated by the subjective choice of baseline when fitting the continuum level interactively in the IRAF software; the intrinsic resolution of the spectrograph is $R \approx 4000$. We note that there is no clear evidence for [\ion{N}{ii}]~$\lambda$6583 or [\ion{S}{ii}]~$\lambda\lambda$6716,31 emission.

Archival VLA data at $8\,$GHz (project ID AM0593) allow us to determine that the core ($S_{\rm 8.4} = 12.6\,\pm\,0.2\,$mJy) is a flat-spectrum object with $\alpha = -0.10 \pm 0.01$. Table \ref{tab:1} lists the possible components of \rgzwat\ and the corresponding NVSS and FIRST identifications.  Using radio components S2 to N3, we estimate the luminosity density to be $L_{\rm 1.4}=5.5\times10^{24}$ W Hz$^{-1}$.  This places \rgzwat\,below the FR\,I/II boundary in a radio-versus-optical luminosity diagram like Fig.\,4 of \citet{Best2009}.  However, \rgzwat\,is still inside the rectangular area where FR\,Is and FR\,IIs occur with almost equal frequency, making it an analogue of local FRI radio galaxies like 3C 31. The lack of strong lines in the spectrum suggests that \rgzwat\,is a low-excitation radio galaxy (LERG).

The Northern section of the radio complex is marked with the labels N1, N2, N3, and N.E. bridge in Fig.~\ref{fig:2}.  In Fig.~\ref{fig:4}(a) we show the diffuse emission connecting components N2 to N3.  The N.E. bridge is detected in the lower-resolution NVSS image at a peak brightness of $\approx4$\,mJy\,beam$^{-1}$ with a total flux density of 23~mJy spread over 5 NVSS beams, and is not detected in FIRST.  We note that N3 may originate from a faint galaxy SDSS J082328.28+033733.2 (Fig.~\ref{fig:4}b) at a spectroscopic redshift of $z = 0.2601$ \citep{Adelman2006}. Deeper radio observations are required to determine if component N3 is connected to the rest of \rgzwat. 

\begin{figure*}
	\includegraphics[scale=0.8]{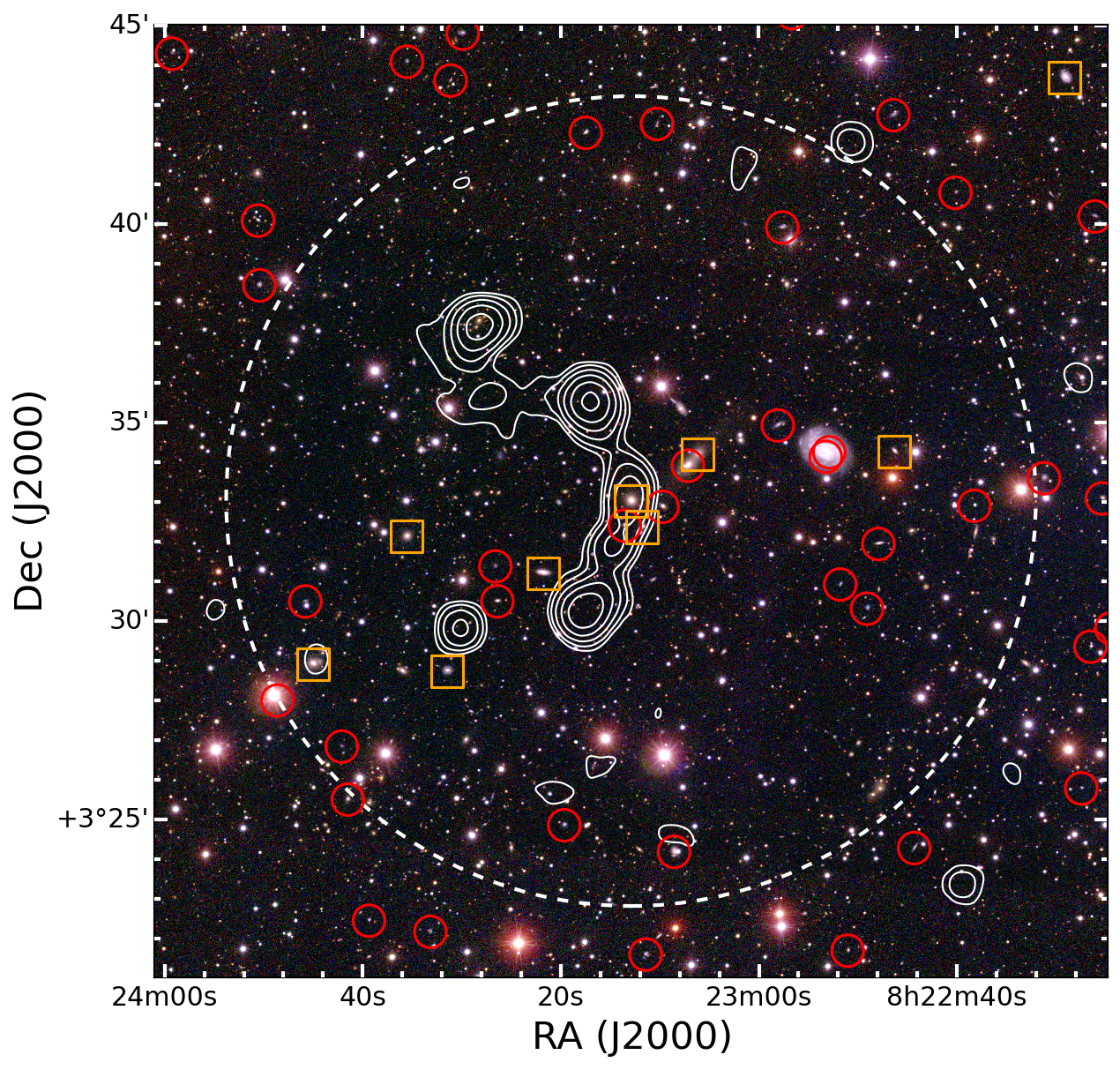}
    \caption{A colour composite of the environment within $1\,$Mpc (10
    arcmin, the radius of the white dashed circle) of \rgzwat.  The
    background is the $rgb$ image of the SDSS field using the $i$ band
    image as red (min $=-0.002$, max $=0.540$), green as the $r$ band
    (min $=-0.006$, max $=0.710$) and the $g$ band for blue (min
    $=-0.002$, max $=0.280)$, all using the asinh
    stretch \citep{Lupton2004}.  The NVSS 1.4-GHz radio flux density is shown with the white contours as displayed in Fig.~\ref{fig:2}.  The orange squares indicate the galaxies with a spectroscopic redshift of $0.08 < z < 0.09$ and the red circles indicate the galaxies with a photometric redshift of $0.05 < z < 0.12$. A colour version of the figure is available in the online version.}
    \label{fig:5}
\end{figure*}

Given the presently available data, there is no hint of diffuse emission connecting components S2 and S3. However, preliminary analysis of the radio structure from DnC configuration Karl G. Jansky Very Large Array observations (Heywood et al. in preparation) indicate a diffuse radio structure to the south of S2 as marginally detected in the NVSS data (Fig.~\ref{fig:5}).  This structure is not included in the analysis of this current work.  The WAT has a projected largest angular size of LAS $\approx 8\,$ arcmin, corresponding to $800\,$kpc, and the total length along the curved ridge of jets/trails is $\approx 1.1\,$Mpc in projection. This makes this WAT comparable in size to 4C+47.51, which, to our knowledge, is still the largest WAT reported by \citet{Pinkney1994}.

\section{Discussion}\label{sec:dis}


\subsection{Environment of RGZ J082312+033301}\label{sec:enivron}
\begin{table*}
 \centering
  \caption{The 23 galaxies with a spectroscopic redshift of $0.08 < z
  < 0.09$ within $31$ arcmin of \rgzwat.  The table includes the name
  of the object, position, SDSS $r'$ band magnitude, spectroscopic, and the distance in arcmin
  between \rgzwat\ and the object.  The values are obtained from NED and SDSS DR12 \citep{Alam2015}.}\label{tab:2}
 \begin{tabular}{lllccc}
 \hline
  Name & R.A. (J2000) & Dec. (J2000) & $r'$ & $z_{\rm spec}$ & Separation\\
 & (degrees) & (degrees) & (mag) & & (arcmin) \\
 \hline
\rgzwat                  & 125.80381 & +03.55038 & 15.66$\,\pm\,$0.01 & 0.08970$\,\pm\,$0.00020 &  0.0\\
SDSS J082311.78+033222.1 & 125.79912 & +03.53947 & 17.24$\,\pm\,$0.01 & 0.08607$\,\pm\,$0.00002 &  0.7\\
SDSS J082306.16+033412.1 & 125.77570 & +03.57004 & 16.37$\,\pm\,$0.01 & 0.08574$\,\pm\,$0.00003 &  2.1\\
SDSS J082321.81+033112.4 & 125.84089 & +03.52012 & 15.96$\,\pm\,$0.01 & 0.08390$\,\pm\,$0.00002 &  2.9\\
SDSS J082335.54+033207.6 & 125.89813 & +03.53546 & 16.58$\,\pm\,$0.01 & 0.08693$\,\pm\,$0.00003 &  5.7\\
SDSS J082331.46+032844.4 & 125.88110 & +03.47901 & 16.99$\,\pm\,$0.01 & 0.08762$\,\pm\,$0.00001 &  6.3\\
SDSS J082246.34+033416.3 & 125.69312 & +03.57122 & 19.75$\,\pm\,$0.02 & 0.08430$\,\pm\,$0.00003 &  6.7\\
SDSS J082345.00+032855.2 & 125.93754 & +03.48201 & 16.08$\,\pm\,$0.01 & 0.08710$\,\pm\,$0.00002 &  9.0\\
SDSS J082229.09+034341.8 & 125.62124 & +03.72828 & 16.26$\,\pm\,$0.01 & 0.08549$\,\pm\,$0.00002 &  15.3\\
SDSS J082405.88+034419.0 & 126.02453 & +03.73861 & 17.74$\,\pm\,$0.01 & 0.08507$\,\pm\,$0.00001 &  17.4\\
SDSS J082403.25+034601.9 & 126.01360 & +03.76720 & 17.60$\,\pm\,$0.01 & 0.08474$\,\pm\,$0.00001 &  18.1\\
SDSS J082423.97+033958.7 & 126.09989 & +03.66631 & 16.42$\,\pm\,$0.01 & 0.08583$\,\pm\,$0.00002 &  19.0\\
SDSS J082205.11+032332.8 & 125.52133 & +03.39244 & 17.14$\,\pm\,$0.01 & 0.08589$\,\pm\,$0.00003 &  19.4\\
SDSS J082431.70+032859.8 & 126.13211 & +03.48329 & 17.15$\,\pm\,$0.01 & 0.08625$\,\pm\,$0.00001 &  20.1\\
SDSS J082221.02+035201.4 & 125.58760 & +03.86707 & 17.81$\,\pm\,$0.01 & 0.08710$\,\pm\,$0.00002 &  23.0\\
SDSS J082201.96+031805.8 & 125.50822 & +03.30161 & 16.25$\,\pm\,$0.01 & 0.08511$\,\pm\,$0.00002 &  23.2\\
SDSS J082207.20+031422.1 & 125.53001 & +03.23947 & 16.75$\,\pm\,$0.01 & 0.08471$\,\pm\,$0.00002 &  24.8\\
SDSS J082319.91+035915.0 & 125.83297 & +03.98753 & 16.63$\,\pm\,$0.01 & 0.08530$\,\pm\,$0.00001 &  26.3\\
SDSS J082458.64+032903.3 & 126.24436 & +03.48428 & 15.86$\,\pm\,$0.01 & 0.08570$\,\pm\,$0.00002 &  26.7\\
SDSS J082432.04+035246.3 & 126.13351 & +03.87955 & 19.37$\,\pm\,$0.02 & 0.08882$\,\pm\,$0.00002 &  27.9\\
SDSS J082324.85+040213.9 & 125.85357 & +04.03720 & 17.76$\,\pm\,$0.01 & 0.08873$\,\pm\,$0.00001 &  29.4\\
SDSS J082257.82+040227.8 & 125.74096 & +04.04106 & 16.43$\,\pm\,$0.01 & 0.08512$\,\pm\,$0.00002 &  29.7\\
SDSS J082450.69+031331.3 & 126.21122 & +03.22538 & 16.56$\,\pm\,$0.01 & 0.08543$\,\pm\,$0.00001 &  31.2\\
\hline
\end{tabular}
\end{table*}

As radio galaxies of WAT morphology tend to trace rich environments,
in this section we assess the environment of \rgzwat\ using the
optical and X-ray data available to us.

\subsubsection{Optical galaxy counts}

Fig.~\ref{fig:5} shows the SDSS colour composite image of the environment surrounding the host galaxy of \rgzwat.  The dashed circle represents a radius of 1.0 Mpc. Table \ref{tab:2} lists all of the galaxies within 31 arcmin of RGZ J082312+033301 and with spectroscopically measured redshifts from SDSS DR12 \citep{Alam2015} between $0.08 < z < 0.09$. We found no reported galaxies with a spectroscopic redshift between $0.09 < z < 0.10$ within our search radius and none between 0.053 and 0.080 within 20$'$ radius.

To determine the richness of the cluster environment surrounding
RGZJ082312+033301, we use the parameter $N_{1.0}^{-19}$ as described
in \citet{Wing2011}. This is a background subtracted count of all
galaxies brighter than $M_r = -19$ at the redshift of the radio source
and within a radius of $1.0\,$Mpc around the radio source. The
background count rate is determined locally using an annulus centred
on the radio source with a radius from 2.7 to $3.0\,$Mpc. We
find \rgzwat\ to be located in an environment with $N_{1.0}^{-19} =
42\,\pm\,1$. This implies that the cluster environment
surrounding \rgzwat\ is near the poor end of the cluster richness
spectrum; the vast majority of Abell clusters analysed
by \cite{Wing2011} have $N_{1.0}^{-19} > 40$.


However, \citet{Rykoff2014} do not detect a cluster near \rgzwat\ in
their redMaPPer catalog. The galaxy group which we propose as a possible counterpart of radio component N3 was identified with  by \citet{Hao2010} as the brightest galaxy of the cluster GMBCG~J125.86785+03.62590 with $z_{phot}=0.288$, as well as by \citet{Rykoff2014} as the brightest cluster galaxy of redMaPPer cluster RM\,J082328.3+033733.2 with $z_{phot}=0.2647$ and richness $=28\pm3$ (as of v5.10 of the catalogue\footnote{\href{http://risa.stanford.edu/redmapper}{http://risa.stanford.edu/redmapper}}). However, the RGZ WAT host cluster proposed in the present paper is not listed in this cluster catalog, possibly indicating that its richness is $\lambda\lesssim20$.

\subsubsection{Velocity dispersion}

The velocity dispersion of the environment gives an indication of its
virial mass and therefore of its richness. The radial velocity
distribution of the 14 galaxies with spectroscopic redshifts within 20 arcmin (2 Mpc) of \rgzwat\ is
strongly peaked around 25000 km s$^{-1}$. We ran these velocities
through the {\tt ROBUST} software \citep{Beers1990} to estimate the
mean velocity and velocity dispersion, finding $C_{BI} =
25743\,\pm\,100\,$km s$^{-1}$ and $S_{BI} =  373\,\pm\,65\,$km
s$^{-1}$. This velocity dispersion would again be consistent with a
rich group or poor cluster of galaxies (e.g., \citealt{Becker2007}).
Interestingly, with a redshift of $z=0.0897$, \rgzwat\ is an outlier
in the velocity distribution, with a radial velocity of $cz = 26900$
km s$^{-1}$, so that it has a relative (or ``peculiar'') velocity with
respect to the 13 remaining galaxies of $(26900 - 25700)/(1+z) = +1100 \pm 100$
km s$^{-1}$. We return to this point below, Section \ref{sec:dynamics}.

\subsubsection{X-ray emission}

\rgzwat\ lies at the extreme edge of an archival {\it XMM-Newton}
dataset (Project ID 0721900101). At 14 arcmin from the detector
centre, the target is out of the field of view of the two MOS cameras
and barely in the pn field of view. The pn data were moderately
affected by soft proton flaring and were filtered to an exposure time
of 13.8 ks before analysis.

\rgzwat\ is clearly detected as a point-like source in the pn data
coincident with the radio core (C). In the 0.3--0.8\,Kev range there are approximately 40 counts after background subtraction in a 30-arcsec radius centred on the detection, just enough to fit a
rough spectrum on the assumption of a power-law model with a fixed
photon index of 1.8 and Galactic absorption ($N_{\rm H} = 3.77 \times
10^{20}$ cm$^{-2}$, from {\sc colden}). This gives a
background-subtracted 2--10-keV luminosity of $(3 \pm 1) \times
10^{41}$ erg s$^{-1}$, which is entirely consistent with what we
might expect from jet-related emission from the unresolved core, given
its 1.4-GHz flux density \citep{Hardcastle2009}.

There is no visual evidence for additional thermal X-ray emission directly
surrounding \rgzwat , which in itself rules out a rich cluster
environment at this redshift. To make this quantitative we measured
counts in a 60$^\circ$ pie-slice to the NE of \rgzwat\ , excluding the
AGN and extending out to an AGN-centric radius of 280 arcsec (480 kpc). We find a
$3\sigma$ upper limit on the 0.3--8.0-keV counts in this region of
195, leading to a limit on the counts from an assumed circularly
symmetric X-ray environment of $<1170$ counts. The temperature of the
non-detected environment is unknown but we convert this limit to a
luminosity on the assumption of various temperatures in the range $kT
= 1.0$ keV (appropriate for a reasonably rich group) to $kT = 5.0$ keV
(a rich cluster), assuming 0.3 solar abundance and the redshift
of \rgzwat. The bolometric luminosity upper limits implied by this are
between $1.2 \times 10^{43}$ erg s$^{-1}$ for $kT = 1.0$ keV and
$2.6 \times 10^{43}$ erg s$^{-1}$ for $kT=5.0$ keV, which is certainly not
consistent with a rich cluster environment given the well-known
temperature-luminosity relation for groups and clusters. It is,
however, consistent with the measured velocity
dispersion: \cite{Helsdon2000} find that one may expect a luminosity $\sim
10^{43}$ erg s$^{-1}$ for $S_{BI} \approx 400$ km s$^{-1}$. We note
that this is also the typical luminosity for a low-excitation radio
galaxy of this luminosity found in the study of \cite{Ineson2015}.

It is worth noting that there is a marginally significant detection of
extended emission with $90 \pm 30$ background-subtracted 0.3--8.0-keV
counts in a 1-arcmin source circle centred at RA=08h 23m 07.0s,
Dec=+03$^\circ$ 33$'$ 53$''$, 1.7 arcmin (170 kpc) to the NW
of \rgzwat. Given the signal to noise it is impossible to confirm that
this is thermal emission, but it is perfectly possible that it
represents the peak of the thermal emission from a group of galaxies
with the properties estimated above from the optical data, possibly
associated with the bright nearby galaxy SDSS J082306.16+033412.1. The
limits we describe above on emission from around \rgzwat\ would
clearly be consistent with a detection at this level. If so, this
would reinforce the idea that the radio galaxy host is somewhat
dynamically and physically offset from the rest of the environment.
Deeper X-ray data are required to investigate this further. 

We can conclude, based on the optical and X-ray constraints we have, that the environment of \rgzwat\,is consistent with being a rich group or poor cluster of galaxies. We designate this poor cluster as the ``\newcluster\,Cluster'' RGZ-CL J0823.2+0333. Indications of such a cluster have appeared only indirectly in a few references that refer to some of its members as galaxy pairs \citep{Merchan2005,Berlind2006, Wen2009, Keel2013} or to galaxy groups \citep{Tago2006,Tempel2012}.

\subsection{Dynamics of \rgzwat}\label{sec:dynamics}
We want to determine if there is a plausible set of jet and medium
parameters that could explain the large size and bent radio morphology of \rgzwat.
The parameters include the radius of curvature $r_c$, the radio jet
radius $r_r$, density ratios of the radio jets $\rho_r$ to that of the cluster
environment $\rho_{\rm ICM}$, the velocity of the radio jets $v_r$, and the velocity of the galaxy with respect to the cluster's barycentre $v_g$.

The high ratio of the velocity of the WAT host to the velocity
dispersion of the cluster discussed above (1100 km s$^{-1}$ / 373 km
s$^{-1}$ = 2.9, for the galaxies within 20 arcmin) raises the question
of whether it could be a background object, and not bound to the
cluster. However, as a background object it
would not have a significant local thermal plasma to bend the radio
structure. Some perspective on this issue comes from the study of the
dynamical distribution of X-ray AGN in 26 LoCuSS
clusters \citep{Haines2012}. They show that AGN have velocities
between 1 and 3 times the velocity dispersion of their clusters, with
a mean velocity dispersion 1.5 times that of the non-active galaxies.
This distribution is indicative of an infalling population, rather
than a virialized one. The WAT host is consistent with this behaviour,
and might thus be recently encountering the cluster. The spatial
offset between the AGN host and the peak of the X-ray emission, if
real, would also be consistent with such a picture.

Fig.~\ref{fig:2} allows us to place an upper limit on the radius of
the radio jets of $r_r \le 15\,$kpc (9 arcsec) if we use the projected size of the knots or hotspots in the FIRST image.  We can also estimate the radius of curvature of the entire WAT to $r_c = 345\,$kpc by fitting a circle to connect N3 to S2. In order to determine the densities required to produce the observed bending we use a range of density ratios $\rho_r/\rho_{\rm ICM} = 10^{-6} - 10^{-2}$ \citep[e.g., ][]{Douglass2011} based on our observation of \rgzwat\, living in a cluster (Section \ref{sec:enivron}). Using Euler's equation \citep{Jones1979,Begelman1979}:
\begin{equation}\label{eqn}
\frac{\rho_rv^2_r}{r_c} = \frac{\rho_{\rm ICM}v^2_g}{r_r}\, ,
\end{equation}
and the values above we find that the velocity of the radio jets is in the range $0.005{\rm c} < v_r < 0.5{\rm c}$. While this could be suggestive of explaining its bent shape, we note that these are only radial velocities, and to see a significant bend, an appreciable transverse (plane-of-sky) velocity is required.  In addition, if the equation of state is relativistic then the relativistic enthalpy density is more relevant.  The bending depends on the ratio of the jet momentum flux and the ram pressure in the putative cross flow (or, more properly, the pressure change across the jet).  If we know or can estimate the ratio of the pressures in the jet and the ambient medium, we can write the bending formula (Eqn.~\ref{eqn}) in terms of that ratio and the ratio of the jet and cross flow Mach numbers. In this form one finds that NATs require a relatively small Mach number for the jet flow, of typically only a few if the jet and ambient pressures are comparable. In a WAT the jet Mach number would be higher relative to the cross flow Mach number.  Future radio and X-ray observations will provide the necessary data to constrain these parameters.

\subsection{Possible re-ignition of \rgzwat}

 \begin{table}
 \centering
  \caption{The estimated timescale for the possible three episodes of formation.  We use the estimated jet velocity $v_r = 0.05{\rm c}$ for these calculations.}\label{tab:watprop}
 \begin{tabular}{lcc}
 \hline
Comp. & Separation & $t_{\rm min}$\\
 & (kpc) & (Myr) \\
  \hline
C -- N1 & 79 & $5$\\
C -- S1 & 126 & $8$\\
\\
C -- N2 & 294 & $19$\\
C -- S2 & 312 & $20$\\
\\
C -- N3 &  592 & $38$\\
C -- S3 & 538 & $35$\\
\hline
\end{tabular}
\end{table}

We find that \rgzwat\,displays an unusual radio structure extending over a large linear size. Typical features of WAT radio galaxies are regions brightening along the jet trails, especially around the bends. The brightening regions extend farther into bright diffuse components that slowly fade away in brightness. However, \rgzwat\,shows tightly collimated radio structure throughout its extent (Fig.~\ref{fig:2}). Here, we speculate if it is possible that the observed bright regions are hotspots typical of FR~II radio galaxies rather than knots along the jet paths (Fig.~\ref{fig:1}). The existence of hotspots could suggest that the WAT is a re-started (double-double) radio galaxy having had two or three episodes of activity during its lifetime.

Assuming the head jet speed is $v=0.05{\rm c}$ from the velocity range in Section \ref{sec:dynamics}, we evaluate the minimum age of the observed radio structures simply as  $t_{\rm min} = d/v_r$. In Table \ref{tab:1} we list the separation between components and the estimated timescale for $z=0.0897$. We calculate the minimum age of the Northern arm (C\,--\,N2) as $19$~Myr and of the Southern arm (C\,--\,S2) as $20$~Myr.  It typically takes between $0.1\,$Myr and $100\,$Myr for the radio structures to dissipate once the central AGN engine switches off \citep[e.g.~][]{Komissarov1994, Kaiser2000, Kapinska2015}.  With these timescales one would expect hotspots from previous activity to almost entirely fade away leaving behind diffuse emission of the remnant lobes (N2, S2). Component N2 appears to have features consistent with FRII hotspots while component S2 does not, suggesting that the AGN engine has switched off and the hotspots are beginning to dissipate.

Radio galaxies have been shown to undergo multiple active phases separated by periods of quiescence time when the jet production is shutdown \citep[e.g.,~][]{Schoenmakers2000, Kaiser2000, Saikia2009}. The dormant/quiescent phase for the AGN may last between $1000\,$yr and $100\,$Myr \citep[e.g.~][]{Shabala2008, Kunert2011, Shulevski2015}.  The inner components (second phase of AGN activity) may be as young as $5$~Myr (C\,--\,N1) and $8$~Myr (C\,--\,S1).  The implied dormant phase of the order of $10\,$Myr would be consistent with typical timescales for the quiescent phase of radio activity.  Therefore, \rgzwat\,could have had two episodes of AGN activity.

However, the speculative N3 and S3 components would only work in such a scenario if one considers a much rarer case of an inverted triple-double source. A triple-double is a radio galaxy that displays three episodes of activity; at least one such example is known \citep{Brocksopp2007}. An inverted double-double is a re-started radio galaxy in which the hotspots from the new activity episode were formed farther away from the radio core than the previous activity older material. This may happen if the density of the previous activity plasma has decreased enough for the re-started jets to pass easily through. This scenario would require the dormant stage of the order of $100\,$Myr and no hotspots formed within components N1, S1, N2 and S2 \citep[][ and references therein]{Kaiser2000, Marecki2009}. We find no evidence of extended relic radio emission around components N3 and S3, but instead rather compact emission. Therefore, \rgzwat\,does not show characteristics of an inverted triple-double source.

All previously discovered double-double radio galaxies display classical straight radio structures (FR~II morphology). If \rgzwat\,is indeed a re-started radio source, it would pose significant questions as well as constraints on the evolutionary models of re-started radio galaxies \citep[cf.][]{Brocksopp2007, Brocksopp2011}. To resolve this issue both high resolution high radio frequency, and low radio frequency observations are required, which will allow one to investigate the existence of hotspots and determine the spectral ages of the components. We are currently pursuing observations to address these issues and results will be presented in forthcoming publications.

\section{Conclusions}\label{sec:conc}

We presented evidence for  a previously unreported poor cluster of galaxies (\newcluster\,Cluster, or RGZ-CL J0823.2+0333) discovered through an unusual giant WAT radio galaxy found with the citizen science project Radio Galaxy Zoo \citep{Banfield2015}.  The host of \rgzwat\,is also known as 2MASX J08231289+0333016 and has been classified by volunteers of Galaxy Zoo 2 \citep{Willett2013} to be of Hubble type E0. We estimate the $1.4\,$GHz luminosity density to be $L_{\rm 1.4}=5.5\times10^{24}\,$W Hz$^{-1}$.  Using the Oxford Swift IFU spectrograph on the Palomar 5m telescope we found \rgzwat\,to have a redshift $z=0.0897\,\pm\,0.0001$.  At this redshift, we find the largest linear size of \rgzwat\,to be $0.8\,$Mpc, and measuring $1.1\,$Mpc along its curved tails making this giant WAT comparable in size to the largest known WAT 4C+47.51.   

Investigation of the surrounding environment through the $N^{-19}_{1.0}$ measurement by \citet{Wing2011} and through Abell's cluster richness \citep{Abell1989} indicate that \rgzwat\,lives in a cluster near the poor end of the cluster richness spectrum.  

Investigation of the surround environment through the $N_{1.0}^{-19}$ measurement by \citet{Wing2011} indicates that RGZ J082312.9+033301 is located in a cluster near the poor end of the cluster richness spectrum, consistent with Abell's cluster richness classification \citep{Abell1989}. However, RGZ-CL J0823.2+0333 was not found in the redMaPPer catalog \citep{Rykoff2014} and our X-ray analysis indicates that \rgzwat\,lives within a rich group rather than a cluster.  Thus, combining the radio, optical, and X-ray data we conclude that \rgzwat\,lives in an unreported poor cluster of galaxies.

In order to understand the radio morphology of \rgzwat, we have placed limits on the dynamics of the system.  We estimate the velocity of \rgzwat\,to be $v_g \ge 1100\,$km s$^{-1}$ with a jet velocity of $0.005{\rm c} \le v_r \le 0.5{\rm c}$.  Using these values we have shown that \rgzwat\,could have multiple re-triggering episodes of active phases with a dormant phase on the order of $10\,$Myr.  

The discovery of \rgzwat\,shows the benefit of using bent-tail radio sources as beacons of clusters of galaxies. However, the difficulty lies in detecting them in the data of the upcoming radio surveys like EMU, WODAN and MeerKAT MIGHTEE. Expanding on further citizen science projects building on Radio Galaxy Zoo, or on the future development of machine-learning techniques will be key to locating these bent-tail radio sources to find and study clusters of galaxies.

\section*{Acknowledgements}
This publication has been made possible by the participation of more
than 8700 volunteers in the Radio Galaxy Zoo project.  Their
contributions are individually acknowledged
at \href{http://rgzauthors.galaxyzoo.org}{http://rgzauthors.galaxyzoo.org}.
We thank M.\ Chow-Mart\'{i}nez for extracting dynamical parameters
using the {\tt ROBUST} software and Nathan Secrest and Nora Loiseau for information on
the {\it XMM-Newton} data. Parts of this research were conducted by
the Australian Research Council Centre of Excellence for All-sky
Astrophysics (CAASTRO), through project number CE110001020. OIW
acknowledges a Super Science Fellowship from the Australian Research
Council. LR, KW and TWJ acknowledge partial support from the U.S.
National Science Foundation under grant AST-1211595 to the University
of Minnesota. MJH acknowledges support
from the UK's Science and Technology Facilities Council
[ST/M001008/1]. GC acknowledges support from STFC grant ST/K005596/1 and
SV acknowledges an doctoral studentship supported by STFC grant
ST/N504233/1. SS thanks the Australian Research Council for an Early Career Fellowship DE130101399. NS is the recipient of an Australian Research Council Future Fellowship. 

This publication makes use of data products from the {\it Wide-field Infrared Survey Explorer} and the Very Large Array. The {\it Wide-field Infrared Survey Explorer} is a joint project of the University of California, Los Angeles, and the Jet Propulsion Laboratory/California Institute of Technology, funded by the National Aeronautics and Space Administration. The National Radio Astronomy Observatory is a facility of the National Science Foundation operated under cooperative agreement by Associated Universities, Inc.  This research has made use of the NASA/IPAC Extragalactic Database (NED) which is operated by the Jet Propulsion Laboratory, California Institute of Technology, under contract with the National Aeronautics and Space Administration. The figures in this work made use of Astropy, a community-developed core Python package for Astronomy \citep{astropy2013}.


\bibliographystyle{mnras}
\bibliography{mnras-wat.bbl}


\bsp	
\label{lastpage}
\end{document}